\documentclass[a4paper,11pt]{article}
\usepackage{jheppub}

\usepackage{amsmath,amssymb}

\usepackage{subcaption}
\usepackage{CJK}
\usepackage{caption}
\usepackage{float,graphbox}
\usepackage{mathtools}
\usepackage{hyperref}
\usepackage{slashed}
\usepackage[hang]{footmisc}
\newcommand{\Tr}{\mathrm{Tr}}
\newcommand{\beq}{\begin{equation}}
\newcommand{\eeq}{\end{equation}}
\newcommand{\cut}{\mathrm{Cut}}
\newcommand{\CW}{\mathrm{CW}}

\begin{document}
\begin{CJK*}{UTF8}{}
\CJKfamily{gbsn}

\title{Planar loop integrands from cuts in $D$ dimensions}

\author[a]{Fan Zhu(朱凡),}
\emailAdd{zhufan25@gscaep.ac.cn}
\author[b,c]{Song He (何颂),}
\emailAdd{songhe@itp.ac.cn}
\author[c]{Zhenqi Han(韩振琦)}
\emailAdd{hanzhenqi24@mails.ucas.ac.cn}
\affiliation[a]{Graduate School of China Academy of Engineering Physics, No. 10 Xibeiwang East Road, Haidian District, Beijing,
100193, P.R.China}
\affiliation[b]{New Cornerstone Laboratory, Institute of Theoretical Physics, Chinese Academy of Sciences, Beijing 100190, China}
\affiliation[c]{School of Fundamental Physics and Mathematical Sciences, Hangzhou Institute for Advanced Study and ICTP-AP, UCAS, Hangzhou 310024, China}

\abstract{
We present a direct reconstruction formula for planar loop integrands from $D$-dimensional generalized unitarity cuts in any colored theory. The reconstruction combinatorics is separated from the theory-dependent tree amplitudes entering the cuts: for the $L$-loop $n$-point color-ordered amplitude, the integrand is expressed as a sum over admissible non-scaleless scalar graphs dressed by corresponding cuts in $D$ dimensions; the coefficients are given by the universal M\"obius-inversion formula of the refinement poset, or equivalently one minus the Euler characteristics of associated complexes. 

As an application we write down closed-formulas for loop integrands in pure Yang--Mills theory, where the required cuts are generated by gluing $D$-dimensional tree amplitudes and summing over internal gluon states. We also use the two-loop five-point case as a validation, comparing with known integrand data and after integration-by-parts reduction, with known integrated helicity amplitudes. The same framework also produces compact cut-organized data for larger examples, which are included in the accompanying ancillary files. We also describe the corresponding simplification in maximally supersymmetric Yang--Mills theory, where the absence of bubble and triangle subgraphs reduces the relevant cut poset substantially. 
}

\maketitle
\end{CJK*}

\clearpage
\section{Introduction}
A central lesson of modern scattering-amplitudes research is that loop amplitudes are often more naturally characterized by their singularities than by any particular Feynman-diagram expansion. This perspective is especially important for multi-loop amplitudes, which are needed both for precision collider physics and for understanding the structural simplicity of quantum field theory beyond its Lagrangian presentation~\cite{Elvang:2015rqa,Arkani-Hamed:2017mur,Herrmann:2022nkh}. It has led to several complementary developments, including color--kinematics duality and double copy~\cite{Bern:2008qj,Bern:2010ue,Bern:2012uf,Bern:2015ooa,Bern:2018jmv,Bern:2019prr}, the CHY formulation of tree amplitudes~\cite{Cachazo:2013hca,Cachazo:2013iea,Cachazo:2014xea}, and geometric or surface-based descriptions of amplitudes~\cite{Arkani-Hamed:2017mur,Arkani-Hamed:2023lbd,Arkani-Hamed:2023mvg,Arkani-Hamed:2023swr,He:2018okq}. A basic problem common to these approaches is to reconstruct complete loop integrands directly from physical on-shell data.

At loop level, the most established incarnation of this idea is generalized unitarity, in which an amplitude is constrained by products of lower-point on-shell amplitudes evaluated on multiple cuts~\cite{Cutkosky:1960,Bern:1994zx,Bern:1994cg,Britto:2004nc,Bern:2011qt,BernMorgan:1995MassiveLoop,Forde:2007DirectExtraction,Kosower:2011ty,Abreu:2017hqn}. This method has enabled many explicit multi-loop computations in gauge theories, including pure Yang--Mills and QCD applications~\cite{Brandhuber:2005jw, Badger:2015lda,Dunbar:2016cxp,Badger:2016ozq,Abreu:2017PlanarFiveGluon,Badger:2017jhb,Abreu:2018jgq,Abreu:2018zmy,Abreu:2019odu,Dunbar:2020wdh,JinLuo:2019ThreeLoopFourGluon,Kosower:2022bfv,Carrolo:2026PrescriptiveQCD}. Closely related is prescriptive unitarity, where one chooses bases of integrands dual to prescribed contours or cuts~\cite{Bourjaily:2017wjl,Bourjaily:2019iqr,Bourjaily:2019gqu,Bourjaily:2020qca,Bourjaily:2021vyj,Bourjaily:2021ujs,Bourjaily:2021hcp}. Other recent approaches construct planar Yang--Mills loop integrands from off-shell recursion or from the classical equations of motion~\cite{Tao:2025fch,Tao:2026irg}. The question addressed in this paper is similar in spirit but different in organization: once a canonical non-scaleless set of cuts has been identified, can one write the planar integrand directly as a universal linear combination of those cuts, without first introducing an ansatz containing redundant or scaleless sectors~\cite{Bourjaily:2017wjl,Zhang:2012IntegrandLevelReduction,Mastrolia:2012an,Mastrolia:2012PolynomialDivision,Ita:2015SurfaceTerms}?

For planar colored theories this question has a particularly clean answer. A generic momentum-space representation may contain tadpoles, massless bubbles on external legs, and related degenerations that are scaleless after integration. Such terms obscure the physical content of the integrand and complicate cut matching~\cite{ArkaniHamed:2024SurfaceKinematicsYM,Cao:2025mlt,Ossola:2006us}. Recent surface-based formulations suggest a way to avoid this problem from the beginning: amplitudes are organized by curves on surfaces, and in the planar case the corresponding variables separate cut sectors canonically~\cite{Arkani-Hamed:2023lbd,Arkani-Hamed:2023mvg,Arkani-Hamed:2023swr,Arkani-Hamed:2023jry,Arkani-Hamed:2024nhp,Arkani-Hamed:2024vna,Arkani-Hamed:2024yvu,De:2024wsy,ArkaniHamed:2024TheCutEquation,ArkaniHamed:2024SurfaceKinematicsYM}. This turns reconstruction from cuts into a combinatorial inversion problem rather than a basis-fitting problem.

The mechanism is already visible at one loop, where canonical cuts organized by cyclic partitions of external legs lead to an inclusion--exclusion formula~\cite{Cao:2024olg}. Our first result is that this structure extends uniformly to arbitrary loop order. We organize admissible non-scaleless cut configurations into a refinement poset and show that the full planar integrand is obtained by M\"obius inversion~\cite{Rota:1964}.\footnote{For a broader perspective on M\"obius inversion over partially
ordered decompositions, see Ref.~\cite{Jansma:2024qhq}. We thank
Abel Jansma for pointing out this connection.} Equivalently, the coefficient of each canonical cut is one minus the Euler characteristic of a complex associated with the corresponding topology. The novelty is therefore not a new version of generalized unitarity itself, but a canonical non-scaleless organization in which the reconstruction coefficients are universal and the field theory enters only through the cut functions.

The main outputs of the paper are the following.
\begin{enumerate}
\item We give an all-loop formula for planar color-ordered integrands in any colored theory in $D$ dimensions. The formula uses only admissible non-scaleless topologies and reconstructs the planar integrand from canonical cuts with coefficients determined by the refinement poset.

\item We apply the formula to pure Yang--Mills theory. Since the cuts are generated by gluing $D$-dimensional Yang--Mills tree amplitudes and summing over internal gluon states, the result is an explicit integrand construction, not only a formal cut-combination formula. We use the two-loop examples to show the construction in detail, and emphasize the two-loop five-point case as a stringent check against known integrand data~\cite{Cao:2025mlt} and against known integrated helicity amplitudes after integration-by-parts reduction~\cite{Badger:2017jhb}. The same framework also produces larger compact cut-organized examples, including the two-loop six-point and three-loop four-point cases, for which fully expanded cuts are too large to be useful as static ancillary files.

\item We spell out the analogous reduction for maximally supersymmetric Yang--Mills theory. The absence of bubble and triangle subgraphs leaves a much smaller cut poset, so the same reconstruction formula becomes especially compact once the corresponding supersymmetric cut input is supplied.
\end{enumerate}

The distinction between a cut-combination formula and an explicit integrand is important throughout the paper. The universal M\"obius-inversion coefficients are theory independent, while the theory-dependent cut functions are supplied by tree gluing in the pure-Yang--Mills examples considered below. In maximally supersymmetric Yang--Mills theory we emphasize the reduced cut-combination structure and leave a fully automated supersymmetric tree-gluing implementation for future work. The accompanying ancillary files collect pure-Yang--Mills cut-combination data through four loops at four points and, for maximally supersymmetric Yang--Mills theory, reduced cut-combination data through at least five loops at four points. For the larger pure-Yang--Mills examples, the compact data should be regarded as the practical representation from which expanded cuts can be regenerated when needed.

The paper is organized as follows. In section~2 we derive the general reconstruction formula from non-scaleless planar topologies, spanning cuts, and M\"obius inversion. In section~3 we apply the formula to gauge theory: first as a one-loop warm-up, then to explicit pure-Yang--Mills integrands from $D$-dimensional tree gluing, and finally to the reduced posets relevant for maximally supersymmetric Yang--Mills theory. Section~4 summarizes the results and discusses automated cut generation, non-planar extensions, and analogous constructions for uncolored theories.

\section{All-loop planar integrands from cuts by M\"obius inversion}
This section contains the general construction. We first identify the admissible planar topologies, excluding scaleless sectors before any reconstruction is performed. We then state the cut formula and derive it by M\"obius inversion on the refinement poset. The result is a clean separation: the coefficients are universal and geometric, while the theory dependence is entirely contained in the cut functions.

\subsection{Non-scaleless diagrams and Boolean refinement intervals}

In a generic momentum-space representation, tadpoles, massless bubbles on external legs, and related degenerations may appear at intermediate stages, but they are scaleless after integration and should not be treated as part of a canonical planar integrand. Our construction therefore keeps only \emph{non-scaleless} topologies, namely those denominator structures that carry physical cut information and survive as meaningful building blocks of the reconstruction. Accordingly, a $L$-loop $n$-point diagram/topology $\Gamma$ will be called scaleless if the associated $D$-dimensional scalar integral
\[
\int \frac{d^{LD}\ell}{D(\Gamma)},\quad\text{with } D(\Gamma)=\prod_{i\in\Gamma}D_i\,,
\]
is scaleless. In the appendix of~\cite{Cao:2025mlt}, a systematic criterion is given for deciding whether a planar diagram is scaleless once all propagators $D_i$ are expressed in planar variables~\cite{ArkaniHamed:2024SurfaceKinematicsYM}. In the accompanying ancillary \texttt{Mathematica} file, we also provide a practical command \texttt{scalelessQ[poles\_, n\_, L\_]} for testing whether a planar $L$-loop $n$-point diagram is scaleless. 
By restricting from the outset to non-scaleless diagrams, we never need to introduce scaleless sectors into the reconstruction (\textit{i.e.}, no unphysical contributions at intermediate steps), while retaining the cut data needed for the canonical non-scaleless planar integrand.

All diagrams considered here descend from parent diagrams~\cite{Bern:2014kca}, namely cubic graphs with loop momentum flowing through every propagator, which in cut language corresponds to the maximal cut. A rank-$k$ quotient diagram of $\Gamma$ is obtained by contracting exactly $k$ labeled propagators of $\Gamma$. Throughout, propagators are treated as labeled and counted with multiplicity. We define the \emph{minimal spanning diagrams} to be the non-scaleless diagrams for which all proper quotient diagrams are scaleless. Equivalently, they are the minimal non-degenerate elements in the refinement poset. More generally, whenever the interval between two admissible diagrams is complete under refinement, it forms a locally Boolean lattice, so the poset structure directly governs the inclusion--exclusion relations among cuts. 
\begin{figure}[htbp]
    \centering
    \resizebox{0.92\linewidth}{!}
    {\includegraphics{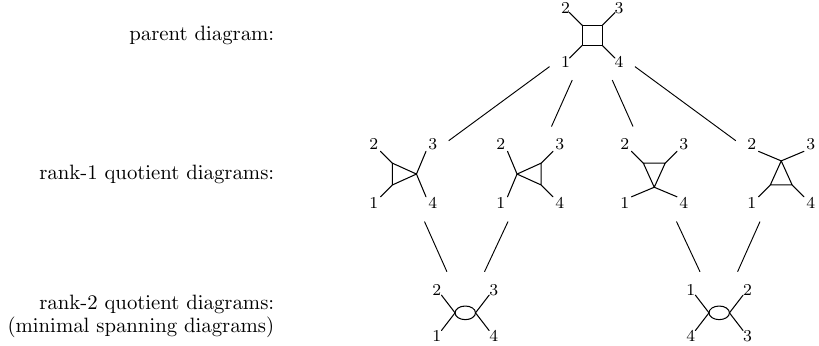}}
    \caption{Refinement poset of planar one-loop four-point diagrams in canonical ordering. The admissible non-scaleless quotient diagrams form a Boolean lattice, with edges corresponding to the contraction of a single propagator.}
    \label{fig_booleanlattice}
\end{figure}

As illustrated in Fig.~\ref{fig_booleanlattice}, the admissible non-scaleless diagrams in this example form a Boolean lattice under refinement. From the viewpoint of generalized unitarity, this means that the cut data are not independent: higher cuts refine lower ones, and the resulting nested relations are exactly those expected from the hierarchy of propagator constraints. More generally, minimal spanning diagrams lie at the bottom of the relevant refinement poset and provide the smallest cut input, while the remaining non-minimal diagrams encode successive refinements. This is precisely the structure for which M\"obius inversion gives the natural solution, allowing one to invert the cut relations and reconstruct the canonical non-scaleless integrand from cut data in a universal way.

\subsection{Direct reconstruction from cuts}

We now state the reconstruction formula for color-ordered planar amplitudes in the leading-color sector. The result takes a particularly simple form when written as a sum over admissible non-scaleless topologies: each topology contributes its canonical cut divided by the corresponding scalar denominator, weighted by a coefficient that depends only on the combinatorics of its descendants in the refinement poset.

\begin{equation}\label{eq_A=cuts}
    A_{n}^{(L)}(\mathbb{I}_{n})=\sum_{\Gamma\in\mathcal{S}_{\mathbb{I}_{n}}^{(L)}}\gamma_{\CW\left(\Gamma\right)}\frac{\cut(\Gamma)}{D(\Gamma)}\,.
\end{equation}
Here $\mathbb{I}_{n}=(1,2,\ldots,n)$ denotes the planar ordering of external legs, $\mathcal{S}_{\mathbb{I}_{n}}^{(L)}$ is the set of admissible non-scaleless planar topologies on a disk with $n$ ordered marked points on the boundary and $L$ punctures in the interior, $D(\Gamma)$ is the product of scalar denominators associated with the diagram $\Gamma$, and $\cut(\Gamma)$ is the corresponding canonical cut. The coefficient $\gamma_{\CW(\Gamma)}$ is one minus the Euler characteristic of the CW complex\footnote{A CW complex (also cellular complex or cell complex) is a topological space that is built by gluing together topological balls (so-called cells) of different dimensions in specific ways.} $\CW(\Gamma)$ associated with $\Gamma$. 
We use the convention
\begin{equation}\label{eq_def_eulerc}
    \gamma=1-\chi,\quad \chi=\sum_{d=0}^{\infty}(-)^{d}\,n_d,\quad n_d\equiv \text{\# of $d$-dimensional cells.}
\end{equation}

For a fixed admissible diagram $\Gamma$, we define $\CW(\Gamma)$ by gluing simplices along faces labeled by the same quotient diagram:
\begin{equation}\label{eq_CW_G}
    \CW(\Gamma)
    =\bigcup_{G\in\mathrm{Min}(\Gamma)}
    \mathrm{Simplex}([G,\Gamma))\,.
\end{equation}
Here $\mathrm{Min}(\Gamma)$ denotes the set of minimal spanning quotient diagrams of $\Gamma$, and $\mathrm{Simplex}([G,\Gamma))$ is a $(|\Gamma|-|G|-1)$-dimensional simplex associated with the refinement interval $[G,\Gamma)$, where $|\Gamma|$ denotes the number of propagators in $\Gamma$. Its $k$-dimensional faces are labeled by the rank-$(k+1)$ quotient diagrams of $\Gamma$ contained in $[G,\Gamma)$. Moreover, the diagram labeling a face is a quotient diagram of those labeling its codimension-one boundary faces. Simplices associated with different minimal spanning diagrams are glued whenever their faces carry the same labeled quotient diagram. In this way, $\CW(\Gamma)$ encodes the complete quotient-diagram combinatorics below $\Gamma$.

If $\Gamma^{\rm min}$ is itself a minimal spanning diagram, then
$[\Gamma^{\rm min},\Gamma^{\rm min})=\varnothing$ and $\CW(\Gamma^{\rm min})$ is understood as the empty complex. Accordingly, $\gamma_{\CW(\Gamma^{\rm min})}=1-\chi\bigl(\varnothing\bigr)=1$.

\begin{figure}[htbp]
	\centering
	\resizebox{0.75\linewidth}{!}{%
		\includegraphics{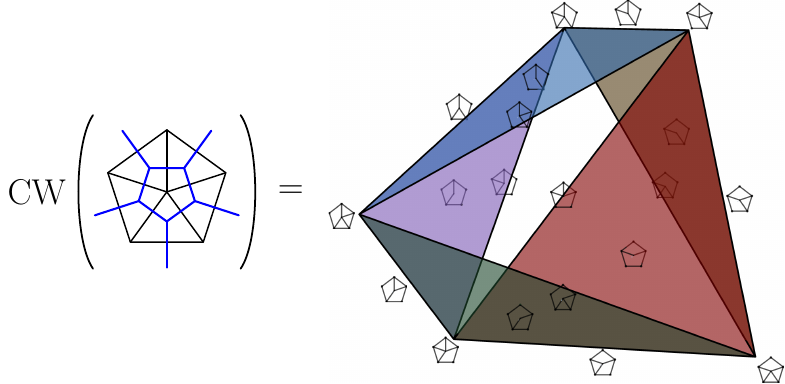}%
	}
	\caption{The CW complex associated with the pentagon diagram of the one-loop five-point amplitude. It is the standard five-vertex triangulation of the M\"obius strip, consisting of five 2-simplices, shown in different colors, together with all their lower-dimensional faces. Each $d$-dimensional face corresponds to a non-scaleless rank-$(d{+}1)$ quotient of the pentagon diagram. In the dual-diagram representation used here, these quotients appear as subdiagrams. The coefficient is $\gamma=1-\chi=1-(5-10+5)=1$.}
	\label{fig_CW_pentagon}
\end{figure}

Figure~\ref{fig_CW_pentagon} shows the associated CW complex for the one-loop five-point pentagon. This example makes concrete how the geometry of quotient diagrams controls the final coefficient. In general, $\CW(\Gamma)$ need not be connected; disconnected complexes simply encode the fact that different non-scaleless degenerations can organize into several disconnected sectors of the refinement poset. In the accompanying ancillary \texttt{Mathematica} file, we provide the practical command \texttt{oneMinusEulerChar[poles\_, n\_, L\_]} for computing the coefficient $\gamma$ of any planar $L$-loop $n$-point diagram.

\subsection{Derivation based on M\"obius inversion}
In this subsection, we derive the reconstruction formula~\eqref{eq_A=cuts} by applying M\"obius inversion to the refinement poset of admissible diagrams.

The M\"obius inversion formula provides a general method for inverting summation relations on partially ordered sets. Although it originated in number theory in the study of arithmetic functions, its natural setting is that of a locally finite poset.

\paragraph{M\"obius inversion on a finite poset.}
Let $(P,\leq)$ be a finite poset with partial order relation ``$\leq$''. 
For functions $f,g:P\to K$, with $K$ a commutative ring, one has
\begin{equation}
	g(x)=\sum_{y\leq x}f(y)
	~~\Longleftrightarrow~~
	f(x)=\sum_{y\leq x}g(y)\,\mu(y,x),\qquad \forall x\in P\,,
\end{equation}
where $\mu(x,y)$ is the M\"obius function defined recursively by
\begin{equation}\label{eq_def_mufunc}
\mu(x,x)=1,\qquad
\mu(x,y)=-\sum_{x\leq z<y}\mu(x,z),\quad (x<y,\ x,y\in P),
\end{equation}
where all sums are finite. On a Boolean refinement interval $[x,y]$, the M\"obius function reduces to an alternating sum:
\begin{equation}
	\mu(x,y)=(-1)^{|y|-|x|}\,.
\end{equation}

\paragraph{Proof of~\eqref{eq_A=cuts}.}
Let $P$ denote the poset of admissible non-scaleless $L$-loop $n$-point diagrams. We equip $P$ with the partial order
\begin{equation}
   \rho\leq\tau\quad\Longleftrightarrow\quad \rho\text{ is a quotient diagram of }\tau.
\end{equation}
The complement of the admissible (non-scaleless) set $P$ is downward closed:
\begin{equation}\label{eq_propercalP}
	\rho\notin P \quad\implies\quad \forall\, \sigma<\rho,\,\, \sigma\notin P\,.
\end{equation}
Consequently, for any $\rho,\sigma\in P$ with $\rho\leq\sigma$, every intermediate diagram also belongs to $P$, namely
\begin{equation}\label{eq_interval_closed}
\rho\leq\tau\leq\sigma,\quad \rho,\sigma\in P
\quad\Longrightarrow\quad
\tau\in P\,,
\end{equation}
otherwise downward closure would imply $\rho\notin P$. Therefore, each interval $[\rho,\sigma]\subseteq P$ is the full refinement interval between $\rho$ and $\sigma$.
We introduce the set of polynomials
\begin{equation}
	U=\{\Delta(\rho)\mid \rho\in P,\ \Delta(\rho)\text{ is independent of propagator variables appearing in }D(\rho)\}\,.
\end{equation}
These polynomials play the role of irreducible numerators associated with admissible diagrams and contain all dependence on the specific colored theory. The integrand is then written as
\begin{equation}\label{eq_IXUP}
A_n^{(L)}(U)=\sum_{\rho\in P}\Delta(\rho)/D(\rho)\,.
\end{equation}
Thus, for a given theory, the entire theory dependence is encoded in the numerator data $U$. For example, a four-point tree amplitude takes the form
\begin{equation}
\begin{aligned}
    &A_4^{(0)}(\{\Delta(s\text{-channel}),\Delta(t\text{-channel}),\Delta(\text{contact})\})\\[5pt]
    =&\Delta(s\text{-channel})/s+\Delta(t\text{-channel})/t+\Delta(\text{contact})/1\,.
\end{aligned}
\end{equation}
The property~\eqref{eq_propercalP} reflects the physical statement that any further degeneration of a scaleless denominator structure remains scaleless. Together with~\eqref{eq_interval_closed}, it ensures that admissible diagrams form complete refinement intervals. 

We now define the canonical cuts of~\eqref{eq_IXUP} by~\footnote{For further details on the relations between irreducible numerators and cuts, refer to~\cite{Ochirov:2016ewn}.}
\begin{equation}\label{eq_def_cut}
\frac{\cut(\rho)}{D(\rho)}
:=\frac{1}{D(\rho)}\underset{D_{i\in\rho}=0}{\mathrm{Res}}\,A_n^{(L)}(U)
=\sum_{\tau\geq\rho}\Delta(\tau)/D(\tau)\,.
\end{equation}
Since the set $\{\frac{\cut(\rho)}{D(\rho)}\mid \rho\in P\}$ is overcomplete for $A_n^{(L)}(U)$, we may expand the integrand as
\begin{equation}\label{eq_FIeq1}
\sum_{\rho\in P}\gamma_{\CW(\rho)}\,\frac{\cut(\rho)}{D(\rho)}=A_n^{(L)}(U)\,.
\end{equation}

To determine the coefficients $\gamma_{\CW(\rho)}$, fix $\sigma\in P$ and compare the coefficient of $\Delta(\sigma)/D(\sigma)$ on both sides of~\eqref{eq_FIeq1}. Since $\Delta(\sigma)/D(\sigma)$ appears in $\cut(\rho)/D(\rho)$ precisely when $\rho\leq\sigma$, we obtain
\begin{equation}\label{eq_FIeq2}
\sum_{\rho\leq\sigma}\gamma_{\CW(\rho)}=1\,.
\end{equation}
This is exactly the form required for M\"obius inversion on $P$. With
\begin{equation}\label{eq_g=1}
    f(\rho):=\gamma_{\CW(\rho)},\qquad g(\sigma):=1,
\end{equation}
we have $g(\sigma)=\sum_{\rho\leq\sigma}f(\rho)$. M\"obius inversion then gives a universal coefficient for every admissible diagram,
\begin{equation}\label{eq_FIeq3}
\gamma_{\CW(\sigma)}=\sum_{\rho\leq\sigma}\mu(\rho,\sigma)\overset{\eqref{eq_interval_closed}}{=}\sum_{\rho\leq\sigma}(-1)^{|\sigma|-|\rho|}\,,
\end{equation}
where $|\sigma|$ denotes the number of propagators in $\sigma$. Therefore, the reconstruction formula~\eqref{eq_A=cuts} follows by summing canonical cuts over the poset $P$, with coefficients determined by M\"obius inversion.
This is precisely the alternating sum over the refinement interval below $\sigma$, and the equivalence between~\eqref{eq_FIeq3} and~\eqref{eq_def_eulerc} follows immediately. 

\begin{table}[htbp]
	\centering
	\begin{tabular}{c||c|c|c|c}
		$n\backslash L$ & 1 & 2 & 3 & 4 \\
		\hline\hline
		4 & (1,2,3) & (3,12,25) & (9,82,340) & (32,894,5515) \\
		5 & (1,3,5) & (4,32,72) & (17,343,1295) & \text{} \\
		6 & (2,8,11) & (9,102,209) & (39,1328,4427) &  \\
		7 & (2,13,17) & (13,270,524) &  &  \\
		8 & (3,28,33) & (23,735,1309) &  &  
	\end{tabular}
	\caption{Numbers of planar topologies at multiplicity $n$ and loop order $L$. Each entry $(N_{\rm min},N_{\gamma\neq 0},N_{P})$ records respectively the number of minimal spanning topologies, the number of topologies with nonvanishing $\gamma$, and the total number of admissible non-scaleless planar topologies. The data illustrate the sparsity of the reconstruction.}
	\label{table1}
\end{table}
 
In practice, it is sufficient to compute the cuts associated with the minimal spanning diagrams. Indeed, every other canonical cut in the same refinement component can be viewed as a further residue, or refinement, of one of these minimal spanning cuts. Thus the explicit tree-gluing computation can be organized around the spanning cuts, while the higher cuts are generated by taking additional residues and serve to cancel overlaps among different cut configurations. As shown in Table~\ref{table1}, a notable consequence of the reconstruction formula is that many canonical cuts have vanishing $\gamma_{\CW}$, significantly reducing the range of the summation of~\eqref{eq_A=cuts}.

Finally, let us emphasize that for a fixed colored theory, the poset $P$ can in principle be reduced further. Whenever a diagram $\rho{\in} P$ has $\Delta(\rho){=}0$---for example, because $\rho$ contains interaction vertices that are forbidden in the theory---it does not contribute to the amplitude and could therefore be removed from $P$. In practice, however, such a theory-dependent reduction is usually not beneficial. In Yang--Mills theory, for example, we find that it generally turns more $\gamma_{\CW}$ on, reducing the sparsity of the reconstruction. We therefore keep the larger, theory-independent choice of $P$.

\section{Gauge-theory applications: explicit Yang--Mills integrands and supersymmetric reductions}

We now apply the master formula~\eqref{eq_A=cuts} to planar gauge-theory integrands. This section illustrates how the universal M\"obius-inversion coefficients combine with theory-dependent cut input: in pure Yang--Mills theory the cuts are supplied by $D$-dimensional tree gluing, while in maximally supersymmetric Yang--Mills theory the same formula acts on a much smaller reduced poset because bubble and triangle subgraphs are absent.

We begin with the one-loop specialization, where the coefficients can be written in closed form. We then turn to pure Yang--Mills examples, using the two-loop cases for explicit checks and the larger cases to illustrate the size of the cut-organized output. Finally, we record the corresponding simplifications for maximally supersymmetric Yang--Mills theory and discuss the computation of cut input.

\subsection{Warm-up: all-multiplicity one-loop formula}
Before turning to genuinely higher-loop examples, it is useful to record the one-loop specialization of the reconstruction formula. In this case the general coefficient collapses to a simple function of the number of corners and massless corners, making the inclusion--exclusion mechanism completely explicit.

At one loop, each topology may be viewed as a polygon whose corners carry consecutive sets of external legs.  The minimal spanning diagrams are the massive bubbles, since their degenerations are tadpoles. Because of this simplicity, one finds a closed formula for the characteristic of a diagram $\Gamma_{m;\,k}$ with $m$ corners and $k$ massless corners:
\begin{equation}
\gamma_{m;\,k}
=\sum_{i=0}^{m-2}(-1)^i\begin{pmatrix}m\\i\end{pmatrix}-(-1)^{m-2}k=(-1)^{m}\bigl(m-k-1\bigr)\,,
\label{eq:oneloop-cmk}
\end{equation}
where the binomial sum accounts for all quotient diagrams down to bubbles, while the term $(-1)^{m-2}k$ removes the contributions of massless bubbles. Thus
\begin{equation}
A_n^{(1)}(\mathbb I_n)
=
\sum_{\Gamma_{m;\,k}\in\mathcal S_{\mathbb I_n}^{(1)}}
(-1)^{m}\bigl(m-k-1\bigr)
\frac{\cut(\Gamma_{m;\,k})}{D(\Gamma_{m;\,k})} .
\label{eq:oneloop-general}
\end{equation}

One immediately sees that the coefficient of a one-loop diagram vanishes if and only if it has a single massive corner, \textit{i.e.} for $\Gamma_{m;\,m-1}$. For example
\begin{equation}
    \begin{aligned}
	&A_6^{(1)}(\mathbb{I}_6)=\left(\frac{\cut(12|3456)}{D(12|3456)}+5\,\text{cyclic.}\right)+\left(\frac{\cut(123|456)}{D(123|456)}+2\,\text{cyclic.}\right)-\\
    &\left(\left(\frac{\cut(123|45|6)}{D(123|45|6)}+\frac{\cut(123|4|56)}{D(123|4|56)}\right) +5\,\text{cyclic.}\right)-2\left(\frac{\cut(12|34|56)}{D(12|34|56)}+1\,\text{cyclic.}\right)\\
    & +\left(\frac{\cut(12|3|45|6)}{D(12|3|45|6)}+2\,\text{cyclic.}\right)+\left(\frac{\cut(12|34|5|6)}{D(12|34|5|6)}+5\,\text{cyclic.}\right) -\frac{\cut(1|2|3|4|5|6)}{D(1|2|3|4|5|6)}\,,
\end{aligned}
\end{equation}
where the slot notation is used only as a shorthand for corners: for instance, $1|2|3|4|5$ is the five-corner diagram with one external leg at each corner, while $12|345$ groups the adjacent pairs $(1,2)$ and $(3,4,5)$ into two corners. 

For maximally supersymmetric Yang--Mills theory, the no-triangle theorem allows the poset to be reduced further, so the characteristic depends only on the number of corners:
\begin{equation}
\gamma^{\text{MSYM}}_{m}
=\sum_{i=0}^{m-4}(-1)^i\begin{pmatrix}m\\i\end{pmatrix}=(-1)^{m}\begin{pmatrix}m-1\\3\end{pmatrix}\,.
\label{eq:oneloop-MSYM-cm}
\end{equation}
\begin{equation}
	A_n^{(1),\,\text{MSYM}}(\mathbb I_n)=\sum_{\Gamma_m\in\mathcal S_{\mathbb I_n}^{(1)}}\gamma^{\text{MSYM}}_{m}\,\frac{\cut(\Gamma_m)}{D(\Gamma_m)}\,.
\end{equation}
For example
\begin{equation}
    \begin{aligned}
    	A_6^{(1),\,\text{MSYM}}(\mathbb{I}_6)=
        &\left(\left(\frac{\cut(12|34|5|6)}{D(12|34|5|6)}+\frac{\cut(123|4|5|6)}{D(123|4|5|6)}\right)+5\,\text{cyclic.}\right)+\Bigg(\frac{\cut(12|3|45|6)}{D(12|3|45|6)}+\\
        &2\,\text{cyclic.}\Bigg)-4\left(\frac{\cut(12|3|4|5|6)}{D(12|3|4|5|6)}+5\,\text{cyclic.}\right)+10\frac{\cut(1|2|3|4|5|6)}{D(1|2|3|4|5|6)}\,.
\end{aligned}
\end{equation}
This already illustrates how maximal supersymmetry simplifies the reconstruction through the smaller size of the poset. In the next two subsections, this contrast will become even more transparent.

Beyond one loop, the reconstruction becomes genuinely nontrivial and thus offers the first meaningful arena in which to test the master formula. As summarized in Table~\ref{table1}, already the planar four-point two-loop case contains 3 minimal spanning topologies, 12 topologies with nonzero coefficients, and 25 admissible non-scaleless topologies in total. We now turn to pure Yang--Mills theory in the large-$N$ limit, where the reconstruction formula is required and the cut data can be generated explicitly by gluing trees. This makes it the natural setting for exhibiting the all-loop content of the construction in a genuinely non-supersymmetric theory.

\subsection{Pure Yang--Mills integrands from $D$-dimensional tree gluing}

In pure Yang--Mills theory, the symbol $\cut(\Gamma)$ in the master formula denotes the state-summed product of color-ordered tree amplitudes associated with the cut topology $\Gamma$. Once the admissible cut topologies and their coefficients are known, the remaining input is standard: glue $D$-dimensional Yang--Mills tree amplitudes across the cut propagators and sum over internal gluon polarizations. The two-loop examples below give explicit checks of the construction, while the higher-loop examples indicate the practical size of the cut-organized output.

\subsubsection{Two-loop examples}
\begin{figure}[htbp]
\centering
\includegraphics{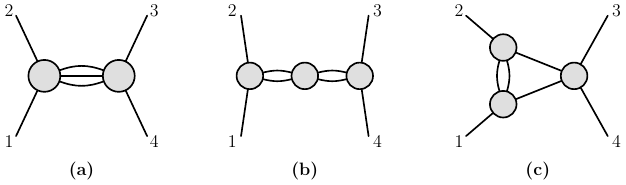}
\caption{Representative minimal spanning diagrams for the three distinct topologies appearing in the two-loop four-point example.}
\label{fig:minseed-three}
\end{figure}

\noindent{\bf 2-loop 4-point:} The simplest genuinely nontrivial example is the two-loop four-point integrand. In this case the canonical input consists of the three minimal spanning cuts shown in Fig.~\ref{fig:minseed-three}.

\begin{figure}[htbp]
	\centering
	\includegraphics[width=0.75\linewidth]{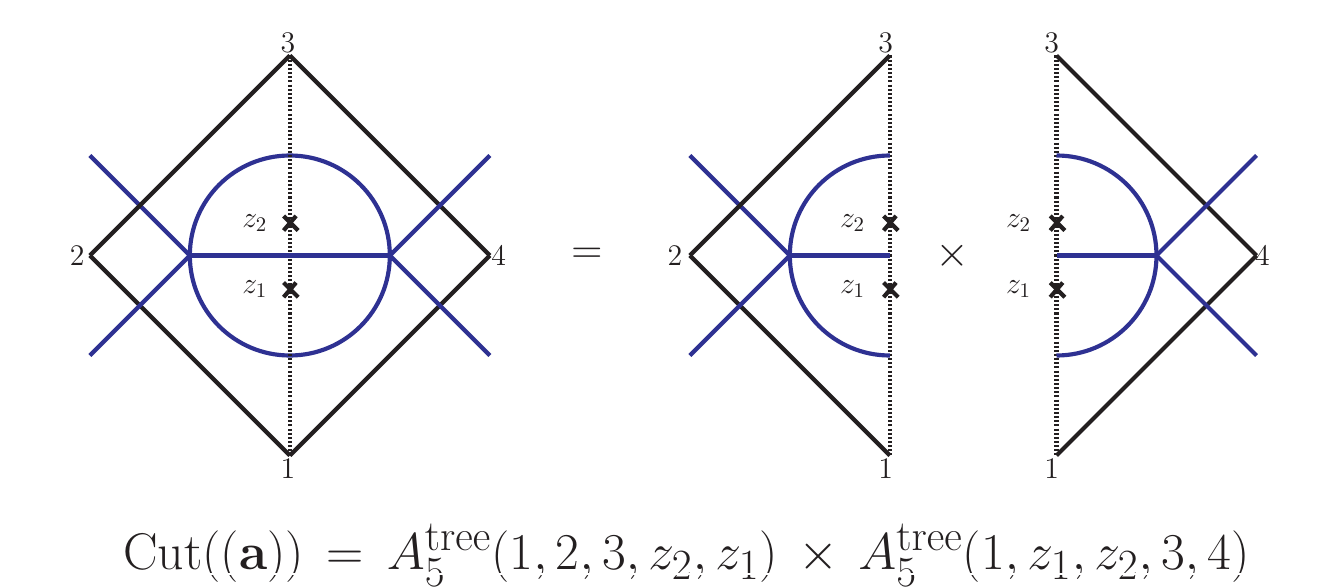}
	\caption{The cut configuration of minimal spanning diagram $\bf (a)$. For planar diagrams, the dual coordinates assigned to each vertex completely determine the momenta entering the corresponding tree amplitude to be glued.}
	\label{fig_cut4_a}
\end{figure}
\begin{figure}[htbp]
	\centering
	\includegraphics[width=1\linewidth]{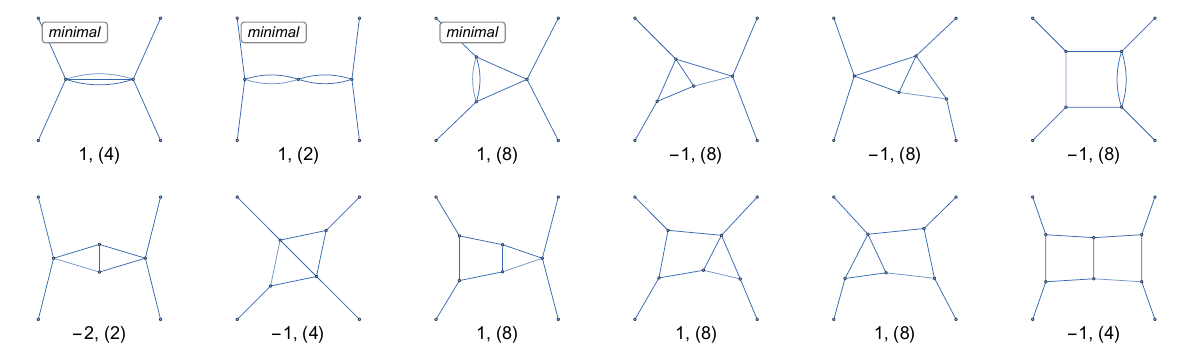}
	\caption{Seed topologies for the planar two-loop four-point integrand, together with their non-zero coefficients $\gamma$ and $C_4\times\mathbb{Z}_2$ orbit sizes $N_{\mathrm{orb}}$. Here $C_4$ acts by cyclic relabelings of the external legs, while $\mathbb{Z}_2$ exchanges the two loop momenta, so
		$N_{\mathrm{orb}}=\bigl|(C_4 \times \mathbb{Z}_2)/\mathrm{Stab}(\Gamma)\bigr|=8/|\mathrm{Stab}(\Gamma)|$.}
	\label{fig_2l4p_seed}
\end{figure}

The three minimal non-scaleless cuts displayed in Fig.~\ref{fig:minseed-three} are the minimal input for the reconstruction. Applying the universal inversion formula generates the full seed representation, whose topologies are shown in Fig.~\ref{fig_2l4p_seed}. The corresponding cut functions are
\begin{equation}\label{eq:2loop4pt-spanning-cuts}
\begin{aligned}
\frac{\cut(\mathbf{(a)})}{D(\mathbf{(a)})}
&=
\frac{
A^{\rm tree}(1,2,3,z_2,z_1)\times
A^{\rm tree}(z_2,3,4,1,z_1)
}{
X_{1,z_1}X_{3,z_2}X_{z_1,z_2}
}\,,
\\[4pt]
\frac{\cut(\mathbf{(b)})}{D(\mathbf{(b)})}
&=
\frac{
A^{\rm tree}(1,2,3,z_1)\times
A^{\rm tree}(1,z_1,3,z_2)\times
A^{\rm tree}(1,z_2,3,4)
}{
X_{1,z_1}X_{1,z_2}X_{3,z_1}X_{3,z_2}
}+(z_1\leftrightarrow z_2)\,,
\\[4pt]
\frac{\cut(\mathbf{(c)})}{D(\mathbf{(c)})}
&=
\frac{
A^{\rm tree}(1,2,z_1,z_2)\times
A^{\rm tree}(2,3,z_2,z_1)\times
A^{\rm tree}(1,z_2,3,4)
}{
X_{1,z_1}X_{2,z_2}X_{3,z_1}X_{z_1,z_2}
}\,,
\end{aligned}
\end{equation}
where the planar variables $X_{\star,\star}$ follow the definitions in~\cite{Cao:2025mlt}. Note that for topology $\bf (b)$, in order to be consistent with the definition~\eqref{eq_def_cut} that $\cut(\Gamma)$ is the residue on $D(\Gamma){=}0$, one must sum over the gluing of two pairs of tree amplitudes, since two distinct topologies share the same propagator structure (these two topologies are related by exchanging the two loop punctures $z_1$ and $z_2$). The binary operation ``$\times$'' denotes gluing tree amplitudes across the cut propagators. In the surface picture, this is equivalently the sewing of adjacent faces, as illustrated in Fig.~\ref{fig_cut4_a}.

\begin{figure}[htbp]
	\centering
	\makebox[\textwidth][c]{%
		\includegraphics[width=1\textwidth]{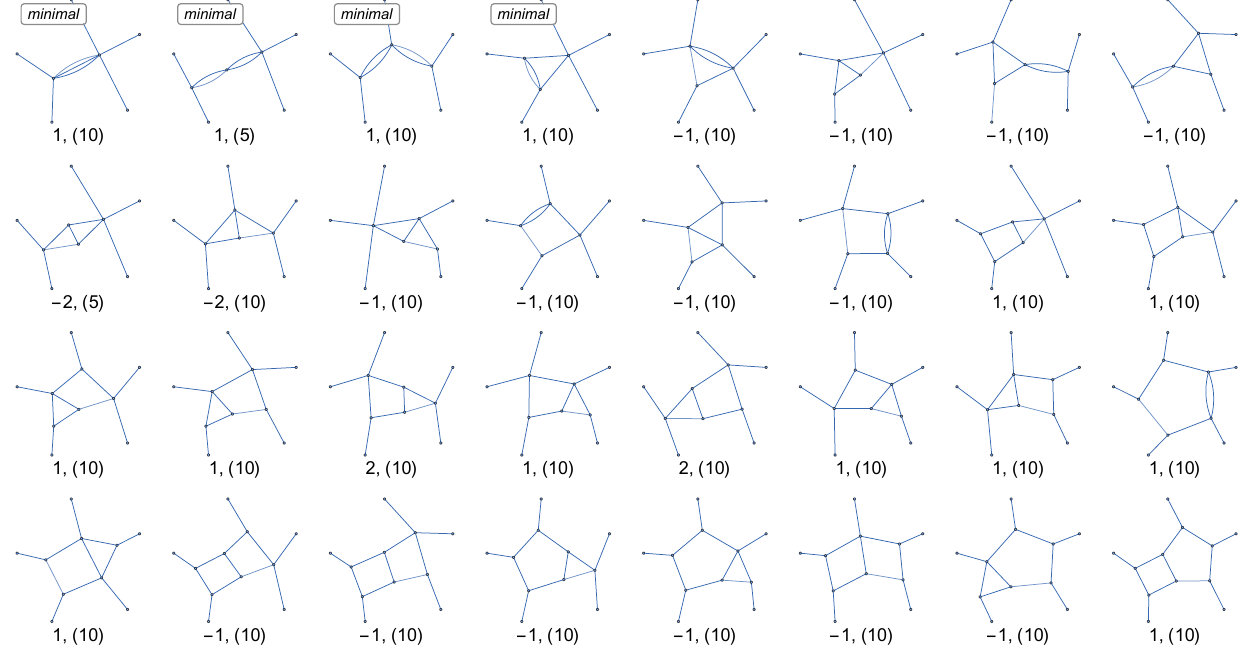}%
	}
	\caption{Seed topologies for the planar two-loop five-point integrand with their non-zero coefficients $\gamma$ and orbit sizes $N_{orb}$.}
	\label{fig:2l5p}
\end{figure}

\vspace{1em}
\noindent{\bf 2-loop 5-point:} The five-point case~\cite{Badger:2015lda} proceeds in the same way but is already much richer. Starting from four minimal spanning cuts, M\"obius inversion produces the 32 seed topologies with nonzero coefficients shown in Fig.~\ref{fig:2l5p}. Substituting the pure-Yang--Mills tree-gluing expressions for these cuts gives a concrete $D$-dimensional planar two-loop five-point integrand in the present canonical non-scaleless representation. This example provides the most stringent check presented here: the result agrees with earlier integrand data~\cite{Cao:2025mlt}, and after integration-by-parts reduction it reproduces the known integrated amplitudes in all independent helicity sectors~\cite{Badger:2017jhb}. Thus the comparison tests the full chain from non-scaleless cut reconstruction and tree-gluing input to the integrated amplitude.

These two-loop examples exhibit the general workflow used throughout the paper: minimal spanning cuts provide the input, M\"obius inversion builds the seed representation, and the simplification is controlled by universal coefficients rather than a theory-specific cancellation. Once the cut functions are supplied by Yang--Mills tree gluing, the output is an explicit integrand representation suitable, in compact cut-organized form, for integration-by-parts reduction and subsequent integration.

\subsubsection{Higher-loop and higher-multiplicity examples}

The same strategy extends to higher loops and higher multiplicity without changing the conceptual input. Here we use the three-loop four-point case mainly to illustrate the practical size of the construction. The three-loop four-point seed representation is displayed in appendix~\ref{app_3l4p}; it involves $82$ topologies with nonzero coefficients, of which $9$ are minimal spanning cuts. The corresponding cut functions can be generated from $D$-dimensional Yang--Mills tree gluing, but their fully expanded form is not a convenient object to print or store. Thus this case is best viewed as compact reconstruction data and as a stress test of the cut organization. At the purely combinatorial level, the same algorithm also produces the cut-combination data displayed in Table~\ref{table1}, including the planar four-loop four-point case, where $894$ topologies have nonzero coefficients and $32$ are minimal spanning cuts.

\subsection{Maximally supersymmetric Yang--Mills theory: reduced cut posets}

For maximally supersymmetric Yang--Mills theory the same reconstruction formula simplifies dramatically. Because cuts obtained by gluing trees vanish whenever a bubble or triangle subgraph is present, the relevant planar topologies form a much smaller subset of the non-scaleless diagrams needed in pure Yang--Mills theory. This makes the supersymmetric case a particularly clean laboratory for exploring the higher-loop structure of the formula. In the present paper we record the corresponding reduced cut-combination data, with ancillary results at least through five loops at four points. A fully systematic implementation of the supersymmetric tree-gluing input is left for future work; the main point here is that once those cuts are supplied, the same M\"obius-inversion formula immediately reconstructs the planar maximally supersymmetric integrand.
\subsubsection{Examples through three loops}

\begin{figure}[htbp]
    \centering
    \setlength{\tabcolsep}{0pt}

    \makebox[\linewidth][c]{%
    \hspace*{-0.02\linewidth}%
    \begin{tabular}{@{}c@{}|@{\hspace{0.02\linewidth}}c@{}}
    	$\mathcal{M}_{4}^{(2)}:
        \vcenter{\hbox{\includegraphics[width=0.16\linewidth]{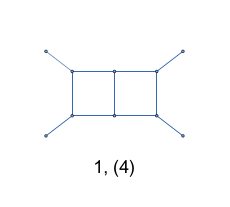}}}$
        &
        $\mathcal{M}_{5}^{(2)}:
        \vcenter{\hbox{\includegraphics[width=0.65\linewidth]{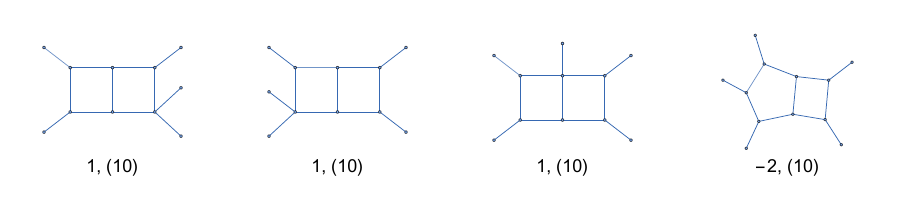}}}$
    \end{tabular}%
    }
    \caption{Seed topologies for 2-loop 4- and 5-point for MSYM with their non-zero coefficients $\gamma$ and orbit sizes $N_{orb}$.}
    \label{fig:susy-2l4p-2l5p}
\end{figure}

\begin{figure}[htbp]
    \centering
    \includegraphics[width=0.65\linewidth]{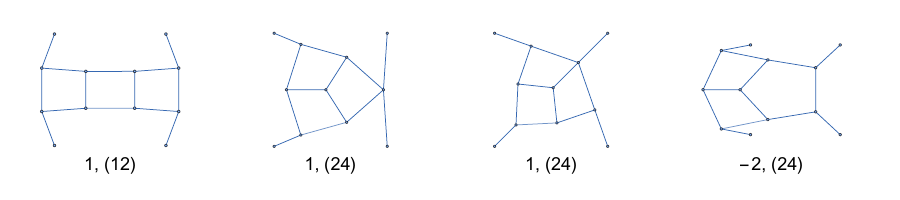}
    \caption{Seed topologies for 3-loop 4-point for MSYM with their non-zero coefficients $\gamma$ and orbit sizes $N_{orb}$.}
    \label{fig:susy_3l4p}
\end{figure}

The first nontrivial examples are shown in Fig.~\ref{fig:susy-2l4p-2l5p} and Fig.~\ref{fig:susy_3l4p}. They illustrate how strongly the reconstruction is simplified once the no-triangle property is imposed on the cut input. In the supersymmetric case, the seed representation is in fact denser, in the sense that a larger fraction of admissible topologies carries a nonvanishing coefficient. The simplification instead comes from the fact that the no-triangle property drastically reduces the total number of admissible diagrams from the outset.

\begin{figure}[htbp]
    \centering
    \makebox[\textwidth][c]{%
    \includegraphics[width=0.9\textwidth]{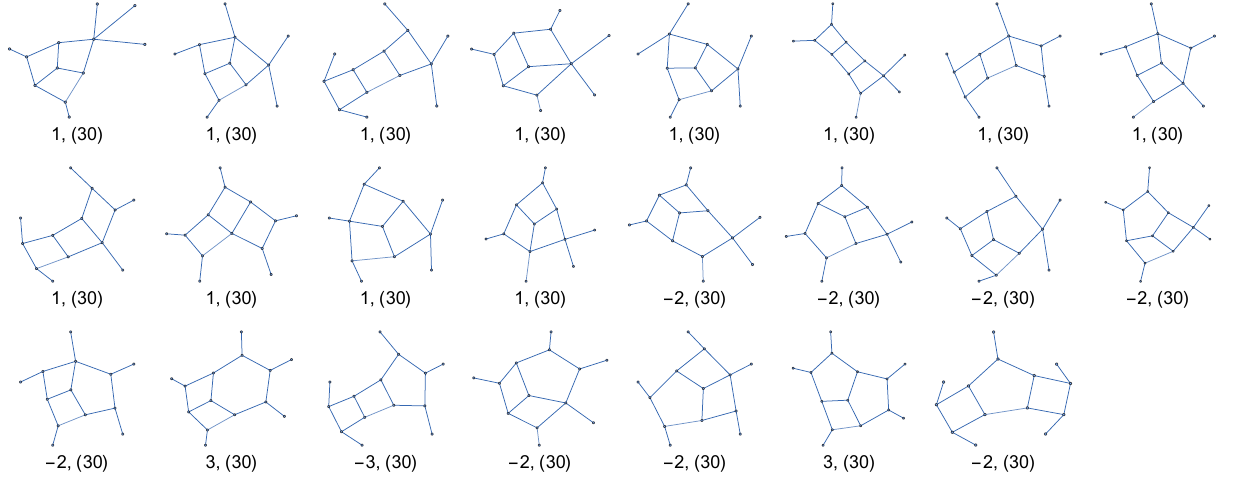}%
    }
    \caption{Seed topologies for 3-loop 5-point for MSYM with their non-zero coefficients $\gamma$ and orbit sizes $N_{orb}$.}
    \label{fig:susy_3l5planar}
\end{figure}

\subsubsection{Four-loop example}
At four loops the reduction is even more striking: only 35 topologies remain in the reduced cut-combination formula for the four-point case. This strongly suggests that maximally supersymmetric Yang--Mills theory is a promising setting for pushing the method to much higher loop order, provided that the supersymmetric cut input can be generated in a comparably systematic way.
\begin{figure}[htbp]
    \centering
    \makebox[\textwidth][c]{%
    \includegraphics[width=0.9\textwidth]{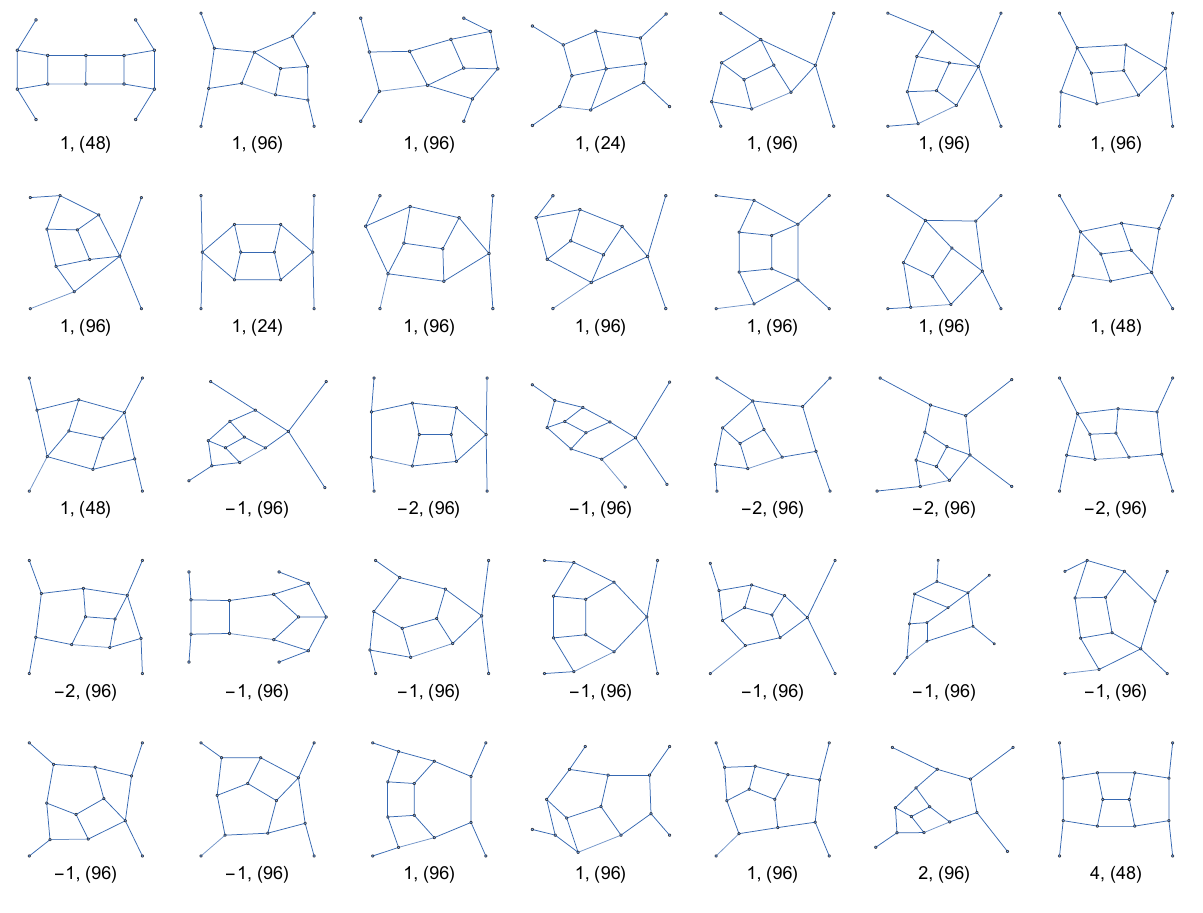}%
    }
    \caption{Seed topologies for 4-loop 4-point for MSYM  with their non-zero coefficients $\gamma$ and orbit sizes $N_{orb}$.}
    \label{fig:susy_4l4p}
\end{figure}

\subsection{Computations of cuts and integration checks}

The examples above separate two tasks that are often intertwined in generalized-unitarity computations. The first task is universal: once the canonical cut functions are known, the passage from cuts to the integrand is fixed by M\"obius inversion. The second task is theory dependent: one must compute the cut functions by gluing the appropriate tree amplitudes and summing over the internal states of the theory.

\paragraph{Pure Yang--Mills cuts.}
For pure Yang--Mills theory, the cut functions are obtained by gluing tree amplitudes in $D$ dimensions~\cite{Cao:2024olg,Edison:2020uzf}. In practice, one first computes the cuts associated with the minimal spanning diagrams. Since every remaining cut in the reconstruction is obtained by imposing additional cut conditions on such a spanning cut, the rest of the cut data can be generated as further residues rather than by starting again from unrelated tree-gluing problems. The internal gluon-state sum is implemented through the standard polarization completeness relation
\begin{equation} \label{eq:statesum}
\sum_{\text{states}} \epsilon_{-}^{\mu}\epsilon_{+}^{\nu}
=\eta^{\mu\nu}-\frac{\ell_{i}^{\mu}q^{\nu}+\ell_{i}^{\nu}q^{\mu}}{\ell_{i}\cdot q},
\qquad
\sum_{\text{states}}\epsilon_{-} \cdot \epsilon_+=D-2,
\end{equation}
where $q$ is a null reference momentum. With this input, the formulae in the previous subsections are explicit pure-Yang--Mills integrands rather than only formal cut identities.

\paragraph{Computational size.}

The main advantage of the present organization is that scaleless integrals are discarded from the outset. Compared with recursive forward-limit constructions~\cite{Cao:2025mlt}, which neither automatically eliminate all scaleless contributions nor avoid generating new ones at intermediate stages, the present approach substantially reduces the amount of unphysical intermediate information. Moreover, those recursive constructions require tree-amplitude building blocks of $(n{+}2L)$-point type, whereas the present inclusion--exclusion strategy only requires $(n{+}L{-}1)$-point tree data. This difference becomes increasingly important at large loop order. The remaining challenge for pure Yang--Mills theory is mostly quantitative: the cut functions themselves can still become very large after tree gluing and state summation. For example, the fully expanded two-loop six-point output contains ${\sim}10^8$ rational terms of Lorentz products. This is why the distinction among minimal spanning cuts, nonzero coefficients, and all admissible non-scaleless topologies is practically important, not merely conceptual. It also explains why the compact cut-organized representation is the sensible input for integration-by-parts reduction: expanding all cuts first would create expressions that are difficult to store, transfer, and reduce.

\paragraph{Integration checks.}
The resulting integrands are intended as starting points for integration-by-parts reduction and integration. In the two-loop five-point case, we have checked the construction against the corresponding integrated amplitudes in all independent helicity sectors after IBP reduction~\cite{Bern:2002tk, Badger:2017jhb}. These checks provide nontrivial validation that the non-scaleless cut reconstruction gives the correct physical amplitudes after integration.

\paragraph{Maximally supersymmetric cuts.}
For maximally supersymmetric Yang--Mills theory, the reconstruction formula is already well suited to much higher orders because the no-triangle property removes many candidate graphs before any cut is evaluated. The corresponding supersymmetric cuts would be obtained by gluing tree-level superamplitudes across the cut propagators and summing over the full on-shell supermultiplet, as in standard generalized-unitarity constructions~\cite{Bern:1994cg,Britto:2004nc,Bern:2011qt,Elvang:2015rqa}. The tree amplitudes can be generated directly in ten-dimensional SYM superspace. Pure-spinor superspace and supersymmetric Berends--Giele currents provide compact ways of organizing the corresponding superamplitudes~\cite{Mafra:2010jq,Mafra:2014oia,Mafra:2015vca}. Equivalently, the state sum may be represented as a Grassmann integration over the cut internal states.

For fermionic states, following the conventions of~\cite{Edison:2022jln}, one may use analogous completeness relations,
\begin{equation}
    \sum_{\rm states}\chi_{-\ell_i}\bar\chi_{\ell_i}
=-\slashed{\ell_i},\qquad
\xi_{-\ell_i}=
\frac{\slashed q\chi_{-\ell_i}}{2q\cdot(-\ell_i)}\,.
\end{equation}
Equivalently,
\begin{equation}
\sum_{\rm states}\bar\chi_{\ell_i}\xi_{-\ell_i}
=
\frac{\Tr(\slashed q\slashed{\ell_i})}{2q\cdot\ell_i}= 2^{D/2-1}.
\end{equation}
Here $\bar\chi_{\ell_i}$ is the spinor wavefunction, $\xi_{-\ell_i}$ is the dual spinor wavefunction, and $q$ is a null reference momentum. The gamma matrices satisfy $\Gamma^\mu \Gamma^\nu + \Gamma^\nu \Gamma^\mu = 2\eta^{\mu \nu}$. We use $\slashed{v}=v^\mu\Gamma_\mu$ and $\slashed{f_i}=\frac{1}{4} f_i^{\mu\nu}\Gamma_\mu\Gamma_\nu=\frac{1}{2}\slashed{k}_i\slashed{\epsilon}_i$.
The ancillary files already contain reduced cut-combination results at least through five loops at four points. A comparably systematic pipeline for generating and organizing the supersymmetric cut functions themselves remains an important next step.

\section{Conclusions and Outlook}

We have constructed planar loop integrands directly from canonical cut data. The key point is to organize the cuts by admissible non-scaleless topologies before performing the reconstruction. Once this is done, the map from cuts to the integrand is a universal M\"obius inversion on the refinement poset~\cite{Rota:1964}. Equivalently, each cut is weighted by one minus the Euler characteristic of an associated complex. This gives a canonical version of generalized unitarity in which scaleless sectors are absent from the outset and the reconstruction coefficients are independent of the underlying field theory.

The main application in this paper is pure Yang--Mills theory in general dimension. In this case the required cut functions are computed explicitly by gluing $D$-dimensional Yang--Mills tree amplitudes and summing over internal gluon states. Thus the formal reconstruction formula becomes a practical method for producing explicit planar pure-Yang--Mills integrands. The two-loop five-point example gives a particularly important check, since after IBP reduction it agrees with the corresponding integrated amplitudes in all independent helicity sectors. At higher multiplicity or loop order, the same construction also makes clear why one should work with compact cut-organized data rather than fully expanded cuts, whose size quickly becomes the main obstacle to direct IBP processing.

Maximally supersymmetric Yang--Mills theory provides a complementary test case. Because bubble and triangle subgraphs are absent, the same reconstruction formula applies to a much smaller cut poset. The reduced cut-combination data already remain compact through relatively high loop orders, suggesting that this theory is a particularly favorable arena for pushing the method further once automated supersymmetric cut generation is implemented.

There are several directions in which the present construction should extend.

\paragraph{Non-planar colored theories.}
The combinatorial logic of cuts, refinements, and M\"obius inversion is not intrinsically planar. Extending the construction beyond the planar limit should therefore be possible, but it requires a comparably canonical definition of loop variables or loop momenta on more general surfaces, together with an appropriate notion of admissible non-scaleless topologies. In the non-planar colored case, one must also decide how to package the color information, either through explicit color factors or through a suitable color decomposition~\cite{DelDuca:1999rs,Bern:2024nonplanar}.

\paragraph{Uncolored theories.}
For uncolored theories such as gravity, the reconstruction step should again be governed by M\"obius inversion, but the cut poset must also keep track of multiplicities associated with equivalent loop-momentum routings. In such a theory one may write schematically
\begin{equation}\label{eq_uncolored1}
	A_n^{(L), \,\text{GR}}
	=
	\sum_{\sigma\in\mathcal{G}}\frac{\Delta(\sigma)}{S(\sigma) D(\sigma)}
	=
	\sum_{\sigma\in\mathcal{G}}\gamma^{\text{GR}}_{\CW(\sigma)}\frac{\cut(\sigma)}{D(\sigma)}\,,
\end{equation}
where $\mathcal{G}$ is a choice of diagrams with assigned loop-momentum routings and $S(\sigma)$ is the number of representatives that become identical to $\sigma$ after integration. For a choice of $\mathcal G$ closed under the relevant refinements, the associated cut data form an overcomplete set for the weighted integrand representation, M\"obius inversion gives
\begin{equation}
    \gamma^{\text{GR}}_{\CW(\sigma)}=\sum_{\rho\leq\sigma}\frac{\mu(\rho,\sigma)}{S(\rho)}\,.
\end{equation}
For example, in the one-loop four-point case, choosing $\mathcal{G}$ to contain three bubble diagrams, twelve triangle diagrams, and twelve box diagrams gives
\begin{equation}
	\begin{aligned}
		&A^{(1),\,\text{GR}}_4=\frac{\Delta(12|34)}{D(12|34)}+\frac{1}{2}\Bigg(\frac{\Delta(1|2|34)}{D(1|2|34)}+\frac{\Delta(2|1|34)}{D(2|1|34)}+\frac{\Delta(12|3|4)}{D(12|3|4)}+\frac{\Delta(12|4|3)}{D(12|4|3)}\Bigg)+\\
		&\frac{1}{4}\Bigg(\frac{\Delta(1|2|3|4)}{D(1|2|3|4)}+\frac{\Delta(1|2|4|3)}{D(1|2|4|3)}+\frac{\Delta(2|1|3|4)}{D(2|1|3|4)}+\frac{\Delta(2|1|4|3)}{D(2|1|4|3)}\Bigg)+(2\leftrightarrow3)+(2\leftrightarrow4)\\
		&=\frac{\cut(12|34)}{D(12|34)}-\frac{1}{2}\Bigg(\frac{\cut(1|2|34)}{D(1|2|34)}+\frac{\cut(2|1|34)}{D(2|1|34)}+\frac{\cut(12|3|4)}{D(12|3|4)}+\frac{\cut(12|4|3)}{D(12|4|3)}\Bigg)+\\
		&\frac{1}{4}\Bigg(\frac{\cut(1|2|3|4)}{D(1|2|3|4)}+\frac{\cut(1|2|4|3)}{D(1|2|4|3)}+\frac{\cut(2|1|3|4)}{D(2|1|3|4)}+\frac{\cut(2|1|4|3)}{D(2|1|4|3)}\Bigg)+(2\leftrightarrow3)+(2\leftrightarrow4)\,.
	\end{aligned}
\end{equation}
Here the loop momentum associated with a one-loop diagram $i|\cdots|e$ is taken to flow from corner $i$ to corner $e$. For color-dressed theories, integrand expansions in irreducible numerators dressed with color factors have already been studied in detail in~\cite{Ochirov:2016ewn}, and a corresponding M\"obius-inversion formulation should be developed in parallel.

\paragraph{Automated cut generation.}
The most immediate computational goal is to automate cut generation more broadly. Pure Yang--Mills theory already demonstrates that the reconstruction formula becomes highly effective once the tree-gluing input is available. Maximally supersymmetric Yang--Mills theory is an especially attractive next target because the reduced poset is small; the remaining task is to implement the supersymmetric tree-gluing data efficiently. More generally, it would be useful to automate cut input for a wider class of gauge theories and to clarify the all-loop families singled out by simple Euler-characteristic patterns.

Overall, the construction suggests that the passage from cuts to amplitudes is a universal combinatorial operation, with the underlying field theory entering only through its tree-level building blocks~\cite{Cachazo:2014nsa,Cachazo:2014xea}. We expect this perspective to be useful both for practical multi-loop computations and for sharpening the relation among generalized unitarity, cut equations, color organization, and surface-based formulations of scattering amplitudes.

\acknowledgments
We are grateful to Qu Cao for collaborations on related projects, and to Johannes Henn, Lorenzo Tancredi and Yang Zhang for encouragement and helpful discussions regarding this direction; we are especially grateful to Xuhang Jiang for carrying out the integration-by-parts reduction and integrated-amplitude checks for the two-loop five-point case. This work is supported by the National Natural Science Foundation of China under Grant No. 12225510 and 12247103, and by the New Cornerstone Science Foundation.

\clearpage

\appendix

\section{Seed topologies for the 3-loop 4-point pure Yang--Mills integrand}\label{app_3l4p}

For the planar three-loop four-point pure-Yang--Mills amplitude, the reconstruction formula gives a compact seed representation involving $82$ topologies, of which $9$ are minimal spanning topologies. The corresponding cuts have been generated by gluing $D$-dimensional Yang--Mills tree amplitudes, but the fully expanded expressions are too large to be useful as static ancillary files. We therefore display the seed topologies here and provide the compact reconstruction data separately. This compact form is the practical input for subsequent manipulations, and the expanded cuts can be regenerated from the code when needed.

\begin{figure}[htbp]
    \centering
    \includegraphics[width=\textwidth]{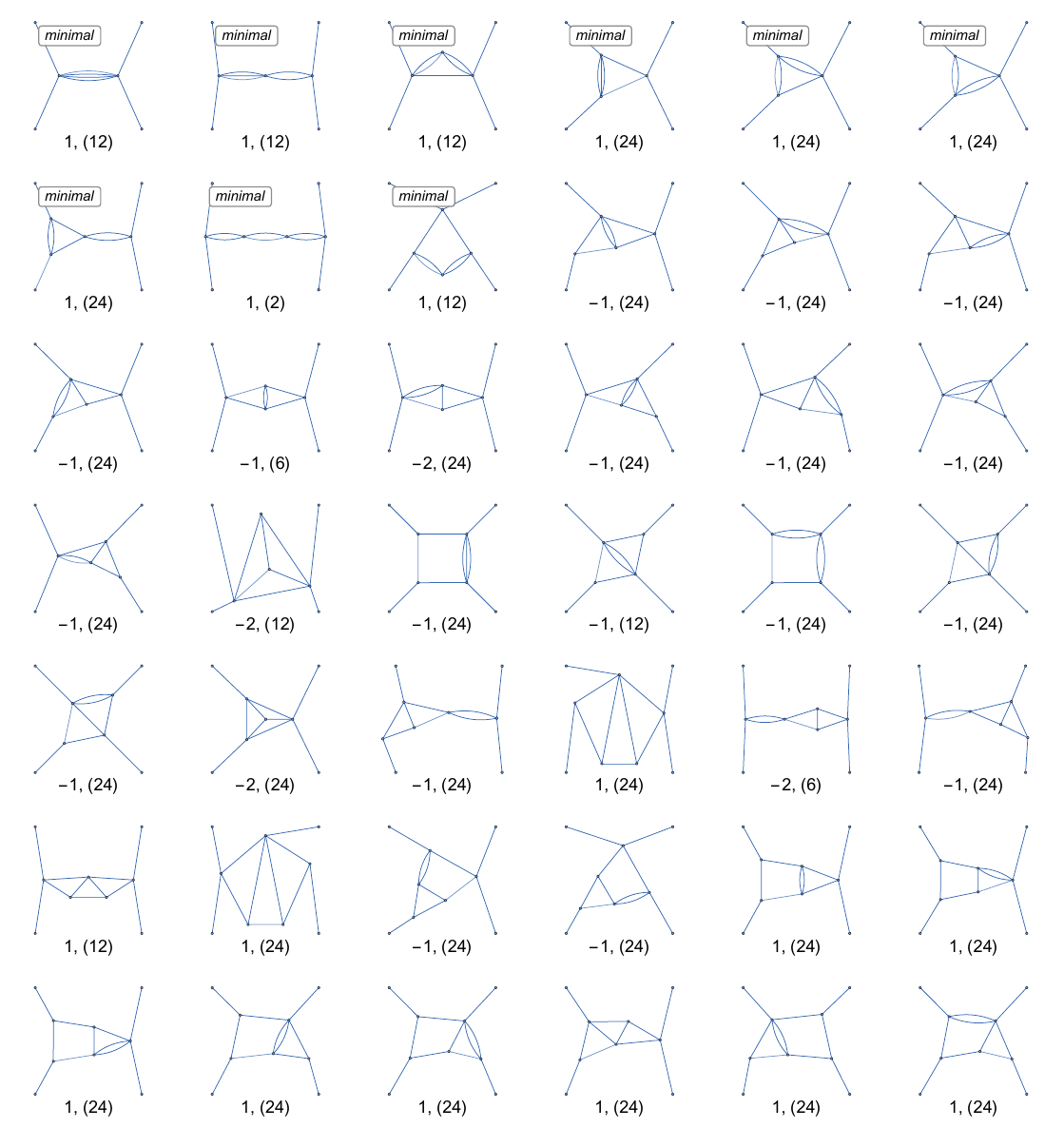}
    \caption{Seed topologies for the three-loop four-point pure-Yang--Mills integrand  with their non-zero coefficients $\gamma$ and orbit sizes $N_{orb}$.}
\end{figure}

\begin{figure}[htbp]
    \ContinuedFloat
    \centering
    \includegraphics[width=\textwidth]{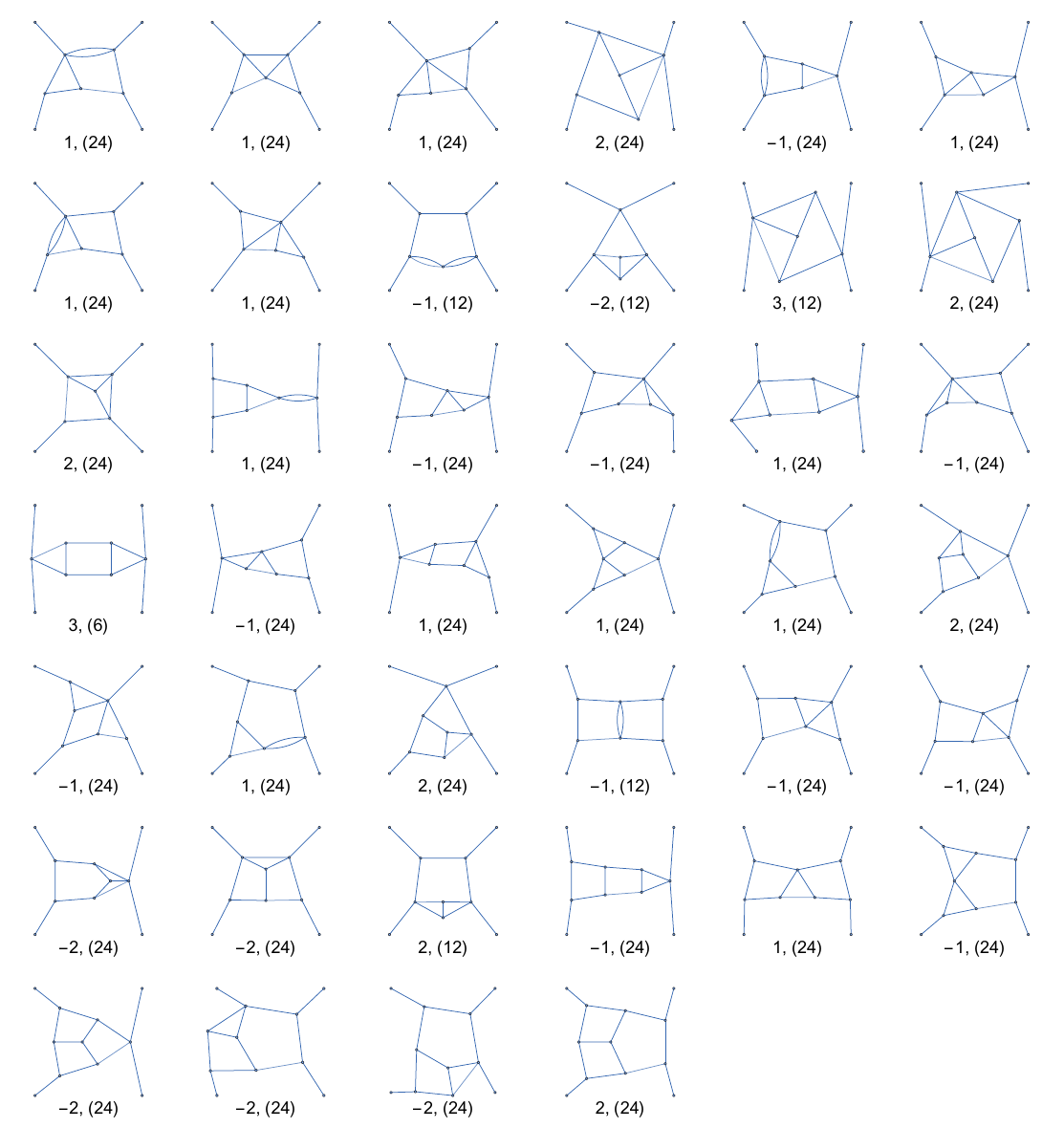}
    \caption{Seed topologies for the three-loop four-point pure-Yang--Mills integrand. (continued)}
\end{figure}

\section{On some all-loop and all-multiplicity patterns}

As summarized in Table~\ref{table1}, the number of admissible non-scaleless topologies grows rapidly with loop order and multiplicity. For sufficiently high loop order or multiplicity, the exhaustive enumeration of all such diagrams becomes one of the main computational bottlenecks of the reconstruction. This makes it particularly valuable to identify all-loop or all-multiplicity patterns that determine in advance which classes of topologies must vanish, or admit simple closed formulas, without requiring a complete diagram-by-diagram analysis. We state several such all-order facts here and include only the proof sketches needed to explain the mechanism.

\begin{figure}[htbp]
	\centering
	\begin{minipage}[c]{0.2\linewidth}
		\includegraphics{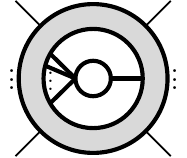}
	\end{minipage}
	\hspace{0.1\linewidth}
	\begin{minipage}[c]{0.2\linewidth}
		\includegraphics{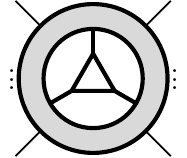}
	\end{minipage}
	\hspace{0.1\linewidth}
	\begin{minipage}[c]{0.2\linewidth}
		\includegraphics{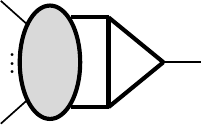}
	\end{minipage}
	\caption{Three representative classes of topologies with vanishing characteristics, associated with internal bubble or triangle configurations. The thick lines denote off-shell propagators.}
	\label{fig_zerotopo}
\end{figure}

From this perspective, families of topologies with vanishing characteristics are especially useful, since they can be discarded \emph{a priori} from the reconstruction. The three basic classes shown in Fig.~\ref{fig_zerotopo}---one containing an internal bubble configuration and two containing internal triangle configurations---have vanishing characteristics. These provide simple and representative mechanisms leading to all-order cancellations.

We do not present the full proofs of these three vanishing theorems here, but only sketch the main idea. Since each non-scaleless quotient diagram contributes $\pm1$ to the reduced Euler characteristic, extensive cancellations occur among the quotient diagrams. A useful organization is to classify the quotient diagrams according to the intersection between the shrunk propagators and the propagators belonging to the bubble or triangle configuration. The resulting pairing among different classes produces sufficient cancellations to make the reduced Euler characteristic vanish.

\begin{figure}[htbp]
	\centering
	\begin{minipage}[c]{0.28\linewidth}
		\centering
		\includegraphics{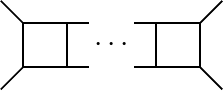}
	\end{minipage}
	$\sim$\hspace{0.02\linewidth}
	\begin{minipage}[c]{0.35\linewidth}
		\centering
		\begin{tabular}{c|c|c|c|c|c|c}
			$L$ & 1 & 2 & 3 & 4 & 5 & $\cdots$ \\
			\hline
			$\gamma_{\text{ladder}}^{(L)}$ & -1 & -1 & 0 & -1 & 1 & $\cdots$
		\end{tabular}
	\end{minipage}
	\caption{The coefficients $\gamma$ of the $L$-loop four-point ladder topologies are signed Fibonacci numbers (\href{https://oeis.org/A039834}{OEIS: A039834}), satisfying $\gamma^{(L+2)}_{\rm ladder}=\gamma^{(L)}_{\rm ladder}-\gamma^{(L+1)}_{\rm ladder}$ with $\gamma^{(1)}_{\rm ladder}=\gamma^{(2)}_{\rm ladder}=-1$.}
	\label{fig_gamma_ladder}
\end{figure}
As a complementary theorem, the ladder family shown in Fig.~\ref{fig_gamma_ladder} has a closed all-loop combinatorial description. Its coefficients are given by signed Fibonacci numbers and satisfy the recursion relation shown in the caption. Simple deformations at the two ends of the ladder, such as adding external legs or shrinking propagators, preserve this structure, changing only the initial conditions of the signed Fibonacci sequence.

Together, these examples illustrate two complementary aspects of the reconstruction: on the one hand, many topologies are eliminated by vanishing characteristics; on the other hand, certain infinite families admit simple closed formulas. Both features suggest that substantial simplifications persist well beyond the low-loop examples displayed explicitly in the main text. 

\clearpage
\bibliographystyle{JHEP}
\bibliography{inspire.bib}

\end{document}